\documentclass[12pt,oneside]{article}
\usepackage{graphicx}

\begin{document}

\begin{center}
  {\large\bf The light quark coupling of the E(38) boson candidate\\}
\vspace{0.14cm}
\vspace*{.05in}
{Chris Austin\footnote{Email: chris@chrisaustin.info}\\
\small 33 Collins Terrace, Maryport, Cumbria CA15 8DL, England\\
}
\end{center}

\begin{center}
{\bf Abstract}
\end{center}

\noindent Assuming that the E(38) boson candidate recently observed at the JINR
Nuclotron is produced in a bremsstrahlung-like manner and decays only to two
photons, its coupling constant to light quarks is estimated to be $\sim
10^{- 4}$.
\vspace{0.65cm}

Recent measurements of $d \left( 3.0 \:
\mathrm{{{GeV}}} / \mathrm{n} \right) +
\mathrm{{{Cu}}} \rightarrow 2 \gamma + X$, $d \left(
2.0 \: \mathrm{{{GeV}}} / \mathrm{n} \right) +
\mathrm{C} \rightarrow 2 \gamma + X$, and $p \left( 4.64 \:
\mathrm{{{GeV}}} \right) + \mathrm{C} \rightarrow 2
\gamma + X$ using the PHOTON-2 electromagnetic lead glass calorimeter at the
JINR Nuclotron \cite{Abraamyan Resonance Structure} have found an excess 
above background in the distribution of
the two-photon invariant mass $m_{\gamma \gamma}$ at about 38 MeV, which has
been interpreted as evidence for the existence of a new light boson, the
E(38), which is not predicted by the Standard Model (SM) {\cite{Abraamyan
E38}}. \ Previous indications for the existence of the E(38) boson were
reported in {\cite{vBR1, vBR2, vBR3}}.

The SM background was measured by the event mixing method, which means that
each pair of background photons consists of two observed photons randomly
selected from different events. \ If the background was pure bremsstrahlung,
and the mean number of bremsstrahlung photons of sufficient energy to 
trigger the calorimeter was the same for all relevant hard SM processes,
and the energies of the photons were negligibly small in comparison to the
momentum transfer between the projectile and the target nucleus, then the
event mixing method would exactly reproduce the true SM background, because
the momenta of bremsstrahlung photons are uncorrelated in this limit. \
Simulation results shown in Fig.\hspace{0.8ex}11 of {\cite{Abraamyan E38}} 
indicate that
the event mixing method provides an adequate approximation to the true
background for the range $10 \: \mathrm{{{MeV}}} <
m_{\gamma \gamma} < 90 \: \mathrm{{{MeV}}}$ relevant
for the search runs, so I shall assume that contamination of the background by
photons from $\pi^0$ decay and other sources is negligible.

I shall use units with $\hbar = c = 1$ and work in the approximation of
treating the energies of the background photons as negligible in comparison to
the momentum transfer between the projectile and the target nucleus. \ The
cuts in practice limit the photon energies to be less than 700 MeV, while the
projectile kinetic energy is not less than 4 GeV. \ The vast majority of
scattering events involve only a small momentum transfer between the
projectile and the target nucleus, but these events do not produce any photons
with sufficient energy to trigger the calorimeter.

I shall call an event ``relevant'' if the momentum transfer between the
projectile and the target nucleus is large enough for the emission of
bremsstrahlung photons of sufficient energy to trigger the calorimeter to be
possible. \ In the Bloch-Nordsieck limit \cite{Bloch Nordsieck}, which I am 
treating as if it were an exact representation of the background, the
number $n$ of bremsstrahlung photons emitted in an event with given initial
and final momenta of all particles other than the bremsstrahlung photons 
follows a Poisson
distribution:
\begin{equation}
  \label{Poisson distribution} P \left( n \right) = \frac{\bar{n}^n
  \mathrm{e}^{- \bar{n}}}{n!},
\end{equation}
where $P \left( n \right)$ is the probability that $n$ bremsstrahlung photons
with energies between $E_{\mathrm{\min}}$ and $E_{\mathrm{\max}}$ are emitted
in the event, and the mean number of bremsstrahlung photons $\bar{n}$ is:
\begin{equation}
  \label{n bar} \bar{n} = - \frac{1}{8 \pi^2} \cdot \mathrm{\ln}
  \frac{E_{\mathrm{\max}}}{E_{\mathrm{\min}}} \cdot \sum_{i j} \eta_i \eta_j
  e_i e_j \frac{1}{\beta_{i j}}\mathrm{\ln} \frac{1 + \beta_{i j}}{1 - \beta_{i
  j}},
\end{equation}
where the sums on $i$ and $j$ run over all electrically charged external
particles of the process, $\eta_i$ is $- 1$ for a particle in the initial
state and $+ 1$ for a particle in the final state, $e_i$ is the electric
charge of external particle $i$, normalized so that if the particle is an
electron, $\frac{e_i^2}{4 \pi}$ is the fine structure constant $\alpha \simeq
\frac{1}{137}$, and $\beta_{i j}$ is the relative velocity of particles $i$
and $j$ in the rest frame of either {\cite{Bloch Nordsieck, Yennie Frautschi
Suura, Weinberg Infrared}}.

From (\ref{n bar}), $\bar{n}$ is 0 if the electrically charged particles and
their velocity vectors are the same in the final state as in the initial
state, and if a set of electrically charged particles whose electric charges
sum to 0 is present both in the initial state and the final state, and their
velocity vectors are equal to one another in the initial state and equal to
one another in the final state, and their final common velocity vector is
equal to their initial common velocity vector, then that set of electrically
charged particles is equivalent to a neutral particle, and can be neglected
for the purpose of calculating $\bar{n}$.

\begin{figure}[t]
\includegraphics[width=0.995\textwidth]{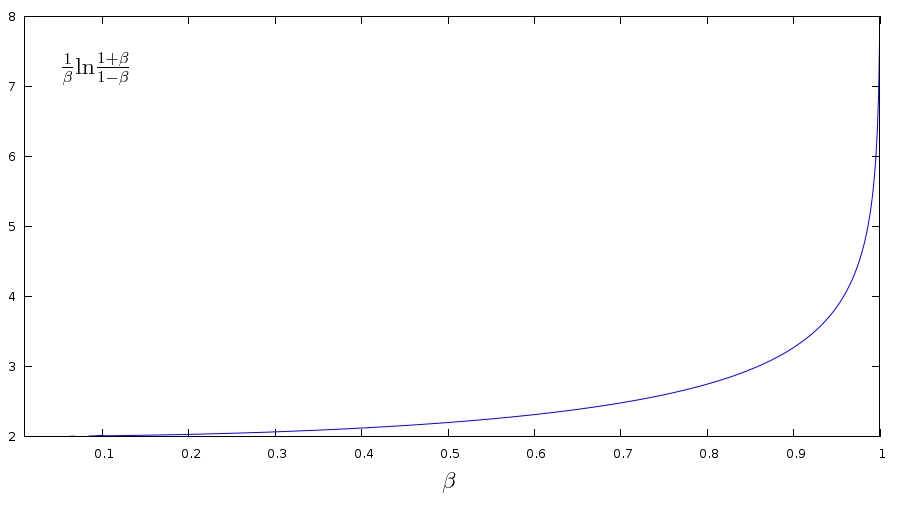}
\caption{The function $\frac{1}{\beta} \mathrm{\ln}
\frac{1 + \beta}{1 - \beta}$, for $0\leq \beta \leq 0.999$}
\label{BN function}
\end{figure}

Figure \ref{BN function} shows that the function $\frac{1}{\beta} \mathrm{\ln}
\frac{1 + \beta}{1 - \beta}$ varies only slowly with $\beta$ for $\beta$ from
0 to about 0.9, and is almost constant for $\beta$ from 0 to about 0.6, so I
shall assume that for a rough estimate of the coupling of the E(38) to the
light quarks, it is sufficient to use the value of $\bar{n}$ for a simple
example of an event whose momentum transfer is high enough for the event to be
relevant. \ I shall use the value of $\bar{n}$ for an event in which the
incident deuteron or proton comes to rest in the target nucleus, while
knocking either two or one neutrons out of the target nucleus so as to
conserve energy and momentum in the limit of exact isospin symmetry.

For this event, the kinetic energies of the incoming projectile and the
ejected neutrons or neutron are more than a GeV and thus much greater than the
kinetic energies of any of the nucleons in the initial and final stationary
nuclei, and $\beta$ for an electron bound to a nucleus of atomic number $Z$ is
not more than about $Z \alpha = \frac{Z}{137}$ in magnitude, so not more than
about 0.04 and 0.2 for the carbon and copper targets respectively, whereas the
values of $\beta$ for the incoming projectile and the outgoing neutrons are
$\beta = 0.971$ for $d \left( 3.0 \: \mathrm{{{GeV}}}
/ \mathrm{n} \right) + \mathrm{{{Cu}}}$, $\beta =
0.948$ for $d \left( 2.0 \: \mathrm{{{GeV}}} /
\mathrm{n} \right) + \mathrm{C}$, and $\beta = 0.986$ for $p \left( 4.64 \:
\mathrm{{{GeV}}} \right) + \mathrm{C}$. \ Thus the
charged nucleons and the electrons of the target atom are effectively at rest
in comparison to the projectile and the outgoing neutrons in both the initial
state and the final state, so to a first approximation the only relevant
charged particle is the projectile. \ So we have:
\begin{equation}
  \label{one charged particle} \bar{n} \simeq - \frac{2 e^2}{8 \pi^2} \cdot
  \mathrm{\ln} \frac{E_{\mathrm{\max}}}{E_{\mathrm{\min}}} \cdot \left( 2 -
  \frac{1}{\beta} \mathrm{\ln} \frac{1 + \beta}{1 - \beta} \right) =
  \frac{\alpha}{\pi} \left( \frac{1}{\beta} \mathrm{\ln} \frac{1 + \beta}{1 -
  \beta} - 2 \right) \mathrm{\ln} \frac{E_{\mathrm{\max}}}{E_{\mathrm{\min}}}
  .
\end{equation}

I shall assume that for the purposes of an order of magnitude estimate, the
angular distribution of the bremsstrahlung photons can be adequately
approximated as isotropic. \ From Table 1 on page 5 of {\cite{Abraamyan
Resonance Structure}}, the total area of the two arms of the PHOTON-2
calorimeter is $0.848 \: \mathrm{{{metre}}}^2$, so the
total area of the right arm of the calorimeter, where the measurements
reported in {\cite{Abraamyan E38}} were recorded, is $0.424 \:
\mathrm{{{metre}}}^2$. \ From page 3 of
{\cite{Abraamyan Resonance Structure}}, the centre of the front surface of the
calorimeter is $3 \: \mathrm{{{metres}}}$ from the
interaction point, so the right arm of the calorimeter covers a fraction
0.00375 of the solid angle around the interaction point.

For simplicity I shall assume that the E(38) boson candidate decays only to
two photons. \ I shall assume the E(38) is produced in a
bremsstrahlung-like manner, and estimate the effective fine structure constant
for the E(38) by using the measurements in {\cite{Abraamyan E38}} to estimate
the ratio of the average number of E(38)'s produced per relevant event to the
number of bremsstrahlung photons with energy between $E_{\mathrm{\min}}$ and
$E_{\mathrm{\max}}$ per relevant event, where $E_{\mathrm{\min}}$ and
$E_{\mathrm{\max}}$ are determined by the cuts applied for each separate plot
in {\cite{Abraamyan E38}}. \ The energy of each E(38) will also be between
$E_{\mathrm{\min}}$ and $E_{\mathrm{\max}}$ up to a possible factor of 2, and
since $\bar{n}$ depends only logarithmically on
$\frac{E_{\mathrm{\max}}}{E_{\mathrm{\min}}}$, I shall use the approximation
of treating the energy of each E(38) as being between $E_{\mathrm{\min}}$ and
$E_{\mathrm{\max}}$.

To test the robustness of the excess above background around $m_{\gamma
\gamma} = 38 \: \mathrm{{{MeV}}}$ against changes in
the choice of cuts and background normalization, 4 different sets of cuts on
the photon energies and the angle between pairs of photons were used, and for
the $d \left( 3.0 \: \mathrm{{{GeV}}} / \mathrm{n}
\right) + \mathrm{{{Cu}}}$ process, the results were
reported for all 4 sets of cuts, and for all 4 of these, results for two
different background normalizations were reported. \ Evidence for the E(38)
boson candidate was found for all choices of the cuts and the background
normalization, but the different sets of cuts are not equally suitable for
estimating the coupling constant of the E(38) to the light quarks, because
they violate by differing amounts the factorization assumption which I shall
use to estimate the single photon background from the reported two photon
background.

The 4 sets of cuts fall
into two groups, called ``soft'' and ``hard''.
For the ``soft'' cut criteria, sets (A) and (B), the angle $\theta_{\gamma
\gamma}$ between the photons is required to satisfy $\mathrm{\cos}
\theta_{\gamma \gamma} < 0.997$, so that $\theta_{\gamma \gamma} >
4.44^{\circ}$, while for the ``hard'' cut criteria, sets (C) and (D),
$\theta_{\gamma \gamma}$ is required to satisfy $\mathrm{\cos} \theta_{\gamma
\gamma} < 0.994$, so that $\theta_{\gamma \gamma} > 6.28^{\circ}$. \ The lower
bound on $\theta_{\gamma \gamma}$ spoils the factorization of the distribution
of the background photons which I shall use to relate the average number of
background photon pairs per relevant event to the average number of single
background photons per relevant event, and the impact of this is more severe
for the ``hard'' cut criteria.

If we approximate the front surface of the right arm of the calorimeter as a
circle of area $0.424 \: \mathrm{{{metre}}}^2$ and
radius $r \simeq 0.37 \: \mathrm{{{metres}}}$, then
the angle between lines from the interaction point to the centre of that
circle and to a point on its perimeter is about $7.1^{\circ}$. \ So for the
``soft'' cut criteria, the lower bound on $\theta_{\gamma \gamma}$ excludes a
fraction 0.39 of the area of that circle for a second photon if the first
photon passes approximately through the centre of the circle, while for the
``hard'' cut criteria, the lower bound on $\theta_{\gamma \gamma}$ excludes a
fraction 0.78 of the area of that circle for the second photon.

I shall therefore use only the measurements made with the ``soft'' cut
criteria (A) and (B), and make the approximation of treating the lower bound
of $4.44^{\circ}$ on $\theta_{\gamma \gamma}$ as if it was 0, so that the
angular distribution of the background photons factorizes.

The cuts on the photon energies also partly violate the requirement that the
distribution of the background photons factorizes, but this is less serious
than for the angle between the photons, because $\bar{n}$ only depends
logarithmically on $E_{\mathrm{\min}}$ and $E_{\mathrm{\max}}$. \ The
factorization requirement is satisfied slightly better for the criteria (A)
than for the criteria (B), because the lower bound on the sum of the photon
energies is lower for criteria (A) than for criteria (B), $E_{\gamma_1} +
E_{\gamma_2} > 300 \: \mathrm{{{MeV}}}$ instead of
$E_{\gamma_1} + E_{\gamma_2} > 350 \:
\mathrm{{{MeV}}}$, so the independent lower bounds
$E_{\gamma_1} > 50 \: \mathrm{{{MeV}}}$, $E_{\gamma_2}
> 50 \: \mathrm{{{MeV}}}$ play a greater role for
criteria (A) than for criteria (B). \ I shall use the results for criteria (A)
to estimate the coupling of the E(38) boson candidate to the light quarks, and
make the approximation of treating the limits on the photon energies for
criteria (A), namely $E_{\gamma} > 50 \:
\mathrm{{{MeV}}}$, $300 \:
\mathrm{{{MeV}}} < E_{\gamma_1} + E_{\gamma_2} < 750
\: \mathrm{{{MeV}}}$, as if they were fully
independent limits on $E_{\gamma_1}$ and $E_{\gamma_2}$. \ The results for
criteria (A) are only presented in {\cite{Abraamyan E38}} for the $d \left(
3.0 \: \mathrm{{{GeV}}} / \mathrm{n} \right) +
\mathrm{{{Cu}}}$ process.

The results for the $d \left( 3.0 \: \mathrm{{{GeV}}}
/ \mathrm{n} \right) + \mathrm{{{Cu}}}$ process with
the cuts (A) are reported for two different choices of the background
normalization. \ In Fig.\hspace{0.8ex}2(a), on page 2 of {\cite{Abraamyan E38}}, 
the background is normalized to the total pair number, and in 
Fig.\hspace{0.8ex}2(b), on page
3, the background is normalized to the number of pairs in the mass range $20
\: \mathrm{{{MeV}}} \leq m_{\gamma \gamma} \leq 28 \:
\mathrm{{{MeV}}}$. \ Reading the results from the graphs, Fig.\hspace{0.8ex}2(a) 
has a total of $159.3 \times 10^3$ background photon pairs,
and $2.32 \times 10^3$ events in the signal excess around 38~MeV. \ So with
the background normalization of Fig.\hspace{0.8ex}2(a), there are $\frac{2.32}{159.3} =
0.015$ E(38) candidates in the right-hand calorimeter for each background
photon pair in the right-hand calorimeter.

For the criteria (A) on the photon energies as above, I shall use the
approximation that $E_{\mathrm{\max}}$ is $750 \:
\mathrm{{{MeV}}} - 50 \:
\mathrm{{{MeV}}} = 700 \:
\mathrm{{{MeV}}}$, and $\mathrm{E_{\mathrm{\min}} = 50
\: \mathrm{{{MeV}}}}$, so $\mathrm{\ln}
\frac{E_{\mathrm{\max}}}{E_{\mathrm{\min}}} = \mathrm{\ln} \frac{700}{50}
\simeq 2.64$. \ So from (\ref{one charged particle}), with $\beta = 0.971$ for
$d \left( 3.0 \: \mathrm{{{GeV}}} / \mathrm{n} \right)
+ \mathrm{{{Cu}}}$, the average number $\bar{n}$ of
bremsstrahlung photons per relevant event, with energies between
$E_{\mathrm{\min}}$ and $E_{\mathrm{\max}}$, is $\bar{n} \simeq 0.014$. \ So
from the paragraph after (\ref{one charged particle}), the average number
$\bar{n}_{\mathrm{r.h.}}$ of bremsstrahlung photons per relevant event that go
into the right-hand calorimeter is $\bar{n}_{\mathrm{r.h.}} \simeq 5.3 \times
10^{- 5}$. \ So by Poisson statistics, the average number
$\bar{n}_{\mathrm{p}}$ of pairs of bremsstrahlung photons per relevant event
in the right-hand calorimeter is $\bar{n}_{\mathrm{p}} \simeq \frac{1}{2}
\bar{n}^2_{\mathrm{r.h.}} \simeq 1.4 \times 10^{- 9}$. \ So from the results
of Fig.\hspace{0.8ex}2(a) in the preceding paragraph, there are $0.015 
\bar{n}_{\mathrm{p}}
\simeq 0.015 \times \frac{1}{2} \bar{n}^2_{\mathrm{r.h.}}$ E(38) candidates
per relevant event in the right-hand calorimeter, so the average number of
E(38) candidates in the right-hand calorimeter per bremsstrahlung photon in
the right-hand calorimeter, which by the isotropy assumption in the paragraph
after (\ref{one charged particle}) is also the average number of E(38)
candidates per bremsstrahlung photon overall, is $0.015 \times \frac{1}{2}
\bar{n}_{\mathrm{r.h.}} \simeq 4.0 \times 10^{- 7}$.

Doing the same counts for Fig.\hspace{0.8ex}2(b), on page 3 of \cite{Abraamyan 
E38}, the number of background
photon pairs is $152.7 \times 10^3$, and the number of events in the excess
around 38 MeV is $3.53 \times 10^3$. \ So for the Fig.\hspace{0.8ex}2(b) 
normalization of
the background, there are 0.023 E(38) candidates per background photon pair. \
This gives the average number of E(38) candidates per bremsstrahlung photon as
$0.023 \times \frac{1}{2} \bar{n}_{\mathrm{r.h.}} \simeq 6.1 \times 10^{- 7}$.

There are indications that the E(38) boson candidate may couple to the quarks
in proportion to their mass {\cite{vBR2}}, so that relatively small $s
\bar{s}$, $c \bar{c}$ and $b \bar{b}$ admixtures in the nucleon could be
significant for the E(38) coupling to the nucleon. \ But for simplicity, I
shall treat the nucleon as containing $u$ and $d$ quarks only. \ So the ``fine
structure constant'' for light quarks coupling to the E(38) is smaller by a
factor $\sim 4 \times 10^{- 7}$ to $\sim 6 \times 10^{- 7}$ than the
electromagnetic fine structure constant. \ So if $g$ is the coupling
constant for the coupling of the E(38) to the light quarks, which would be
the Yukawa coupling constant if the E(38) is a scalar or pseudoscalar,
then $\frac{g^2}{4
\pi} \sim \frac{4 \times 10^{- 7}}{137}$ to $\sim \frac{6 \times 10^{-
7}}{137}$, so $g \sim 1.9 \times 10^{- 4}$ to $\sim 2.3 \times 10^{- 4}$. \
The assumptions and approximations used to obtain this result mean that it is
an order of magnitude estimate at best, so we find:
\begin{equation}
  \label{E38 Yukawa} g \sim 10^{- 4} .
\end{equation}

The value of $g$ could be substantially larger than this if, for example, the 
E(38) was an SM singlet and decayed
mostly to pairs of light sterile neutrinos, so that its branching ratio to two
photons was actually rather small, instead of being 100\%.


\begin{center}
  {\bfseries{Acknowledgements}}
\end{center}

\noindent I would like to thank Eef van Beveren and George Rupp for helpful
comments either on my blog or by email, 
and Tommaso Dorigo for a helpful blog post.

\end{document}